# Fabrication and Test of Pixelated CZT Detectors with Different Pixel Pitches and Thicknesses

Qiang Li, Alfred III Garson, Paul Dowkontt, Jerrad Martin, Matthias Beilicke, Ira Jung, Michael Groza, Arnold Burger, G. De Geronimo, Henric Krawczynski

*Abstract:* The main methods grown Cadmium Zinc Telluride (CZT) crystals with high yield and excellent homogeneity are Modified Horizontal Bridgman (MHB) and High Pressure Bridgman (HPB) processes, respectively. In this contribution, the readout system based on two 32-channel NCI-ASICs for pixelated CZT detector arrays has been developed and tested. The CZT detectors supplied by Orbotech (MHB) and eV products (HPB) are tested by NCI-ASIC readout system. The CZT detectors have an array of 8×8 or 11×11 pixel anodes fabricated on the anode surface with the area up to 2 cm × 2 cm and the thickness of CZT detectors ranges from 0.5 cm to 1 cm. Energy spectra resolution and electron mobility-lifetime products of 8×8 pixels CZT detector with different thicknesses have been investigated.

## I. INTRODUCTION TO PIXELATED CZT DETECTORS

Over the last two decades, the II-VI semiconductor CdZnTe (CZT) has emerged as the material of choice for room temperature detection of hard X-rays and soft γ-rays. The techniques of growing the crystals, the design of the detectors, and the electronics used for reading out the detectors have been improved considerably over the last few years [1-5]. The material finds applications in medical, homeland security, and in astrophysics and particle physics application. The EXIST (Energetic X-ray Imaging Survey Telescope) mission is a proposed all sky survey hard X-ray survey telescope that would use ~14,000 CZT detectors in a coded mask imager with a detector area of 5.5 m² [6]. The EXIST mission would use 0.5×2×2 cm³ CZT detectors each read out with 1024 pixels at a pitch of 0.6 mm. An example of a particle physics experiment is the neutrinoless double beta decay experiment COBRA [7]. A large-scale COBRA experiment would be made of 420 kg of CZT detectors and use either coplanar grid detectors, or pixelated detectors with a pixel pitch of between 150 and 400 microns. The fine pixilization would make it possible to track the electrons from the double beta decays and to distinguish them from alpha backgrounds [8].

CZT detectors have good electron mobility-lifetime products ($\mu_e\tau_e$ ~ $10^{-3}$-$10^{-2}$ $cm^2$ $V^{-1}$), but poor hole mobility-lifetime products ($\mu_h\tau_h$ ~ $5\times10^{-5}$ $cm^2$ $V^{-1}$). All state-of-the-art CZT detectors use "single polarity" readout schemes, where the main information about the energy of the detected radiation is inferred from the anode signals. Coplanar grid detectors, pixelated detectors, and Frisch grid detectors can overcome the severe hole trapping problem and greatly improve the energy spectra resolution of large-volume CZT detectors. For >0.2 cm thick detectors, 662 keV energy resolutions below 1% full width half maximum (FWHM) require to correct the anode signals for the depth of interaction (DOI). The DOI can be measured with the help of the anode-to-cathode charge ratio or with the time offset between the cathode and anode trigger times. Zhang et al. [9] reported a 662 keV energy resolution of 0.76% FWHM for single pixel events from a pixelated detector with 11×11 pixels and a steering grid. The detector size was 1.5×1.5×1 cm³.

In this paper, we report on the test of detectors from the companies Orbotech and eV-products with a low-noise readout system. In Sect. 2 we describe the ASIC based readout system and results from *I-V* measurements. In Sect. 3, we report on the performance of detectors with different pixel pitches and different thicknesses. Sect. 4 gives a summary and an outlook. Throughout the paper, all resolutions are FWHM resolution, and the noise is given in keV-equivalent (for CZT, the average energy to form an electron hole pair is 4.64 eV).

## II. NCI-ASIC READOUT SYSTEM AND *I-V* MEASUREMENT SYSTEM

### A. NCI-ASIC System and Noise Evaluation

Previously we reported on the tests of CZT detectors with discrete Amptek A250 amplifiers and with a 16-channel ASIC [10]. These two systems had electronic noise corresponding to ~1% FWHM at 662 keV. Here we report on first results obtained with a readout system based on the NCI-ASIC developed by Brookhaven National Laboratory and the Naval Research Laboratory [11-15]. The system to control and read out the ASIC was developed at Washington University.

Qiang Li, Alfred III Garson, Paul Dowkontt, Jerrad Martin, Matthias Beilicke and Henric Krawczynski are with the Dept. of Physics and the McDonnell Center for the Space Sciences, Washington University in St. Louis, 1 Brookings Dr., CB 1105, St Louis, MO 63130 (email: qli@physics.wustl.edu)
Ira Jung is with Physik. Inst., Universität Erlangen-Nürnberg, Erlangen, Germany
Michael Groza and Arnold Burger are with Dept. of Physics, Fisk University, 1000 Seventeenth Avenue North, Nashville, TN 37208-3051
G. De Geronimo is with the Instrumentation Division, Brookhaven National Laboratory, Upton, NY 11973 USA

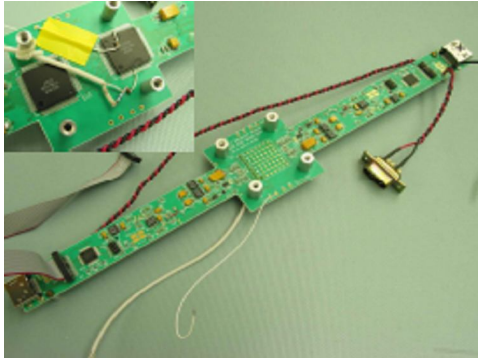

Fig. 1 The readout board developed and used at Washington University for testing CZT detectors with NCI-ASIC developed at Brookhaven National Laboratory and the Naval Research laboratory. The two 32-channel NCI-ASICs are located at the other side of the PC board and can be seen in inserted image.

Fig. 2(a) and Fig. 2(b) show the noise of the NCI-ASIC as function of signal amplitude and as function of detector bias, respectively. All results are obtained with the help of the ASIC's test capacitor. For signals with equivalent amplitudes between 300 keV 1.1 MeV, we obtain an average resolution of 1.3 keV FWHM. The resolution increases to 1.5 keV when a detector is mounted in the system and the detector cathode is biased at -1500 V (resulting in a leakage current of 0.35 nA per pixel).

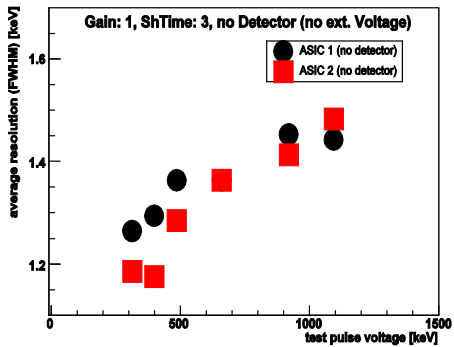

Fig. 2(a) Noise vs. amplitude of input signal.

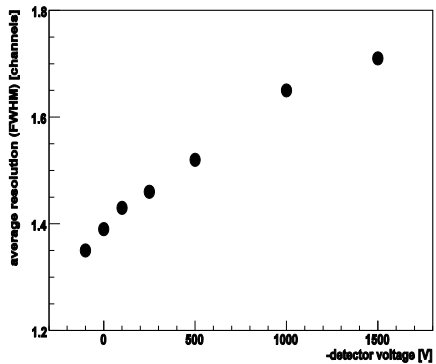

Fig. 2 (b) Noise vs. detector bias.

## B. I-V Measurement System

The leakage current of the CZT detector depends strongly on the contact deposition technique. A system for the automatic acquisition of *I-V* curves of pixilated detectors is shown in Fig. 3(a). The CZT detectors are temporarily mounted in an Ultem plastic holder and are contacted with spring-loaded and gold-plated "pogo-pins". The cathode-pixel, grid-pixel, and pixel-pixel *I-V* curves are automatically measured for detectors with up to 256 pixels using a PC, a computer-controlled Keithley 6514 electrometer, and a programmable high voltage power supply.

A map of the leakage currents of a 1 cm thick eV-Products detector contacted with a Pt cathode and with 64 Au pixels is shown in Fig. 3(b). The maximum leakage current per pixel is ~ 0.5 nA @ -2500 V. The edge pixels have higher leakage current than center ones owing to surface currents along the sides of the detector.

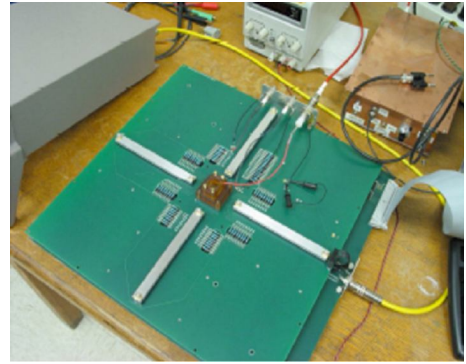

Fig. 3(a) Measurement system for the automatic acquisition of *I-V* curves of pixelated detectors.

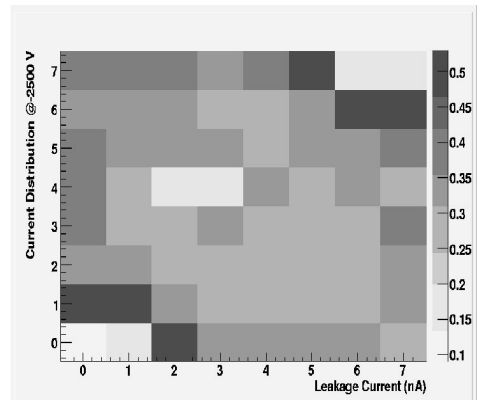

Fig. 3(b) Map of leakage currents of a eV-Product CZT detector ($1\times 2\times 2$ cm$^3$) contacted with 64 pixels. The maximum leakage current per pixel is ~ 0.5 nA @ -2500 V bias voltage.

## III. TEST OF DETECTORS WITH DIFFERENT PIXEL PITCHES AND DIFFERENT THICKNESSES

### A. Experimental Setup

The detector and readout system are operated at room temperature ~22°C. A $^{137}$Cs source is used to flood-illuminate the detector from the cathode side with 662 keV photons. Energy spectra are taken at different cathode biases to enable the measurement of the electron mobility-lifetime products. The detectors are mounted in an Ultem plastic holder, and are connected to a custom built PC board with spring-loaded gold-plated pogo-pins. We contacted the Orbotech detectors with a gold cathode and Ti pixels. The eV-Product detectors have a Pt cathode and Au pixels.

### B. Results obtained with a 1 cm thick 64-pixel detector (2.5 mm pixel pitch)

Fig. 4(a) shows the results from a test of a $1\times2\times2$ cm$^3$ CZT detector from the company eV-Products with a 662 keV $^{137}$Cs source (-1800 V bias voltage, 1 µs shaping time). The detector was contacted with 64 pixels, but only 60 of the 64 pixels were connected to the readout system, as the detector surface of the remaining 4 pixels will be used in future to connect a grid between the pixels to a steering voltage. For this specific detector, we obtained good energy spectra for 59 out of 60 (98%) pixels. NCI-ASICs were used to read out the anode pixels and the cathode, and Fig. 4(b) shows the correlation between the charge induced on one pixel versus the anode-to-cathode charge ratio (only the absolute values of the charges are considered here and in the following). The latter shows a value close to unity for interactions close to the cathode. For interactions closer to the pixels, the absolute value of the charge induced on the cathode decreases more rapidly than the absolute value of the charge induced on the pixels, resulting in a non-linear relationship that can be used to infer the DOI. Fig. 4 (c) shows the energy spectra after correcting the anode signals for the DOI based on the anode-to-cathode charge ratio. The FWHM energy resolution is 0.61% (4.04 keV), the peak to valley ratio is ~30, and the peak-to-Compton ratio is ~10. Averaged over all pixels, the energy resolution of the detector is 0.79% FWHM (5.23 keV).

Fig. 4(d) shows the dependence of the energy resolution (after DOI correction) on the distance of the pixels from the center of the detector. Averaged over pixels at the same or similar distances from the center of the detector, the 662 keV energy resolutions deteriorate from 0.71% FWHM close to the center of the detector to 0.87% FWHM for the edges pixels of the detector anode. As is seen in the figure, the average FWHM of the edge pixels is larger than that of other sections pixels which can be an indication of an unfocused electrical field. The results demonstrate that this CZT detector has not only high spectra grade performance but also good material uniformity.

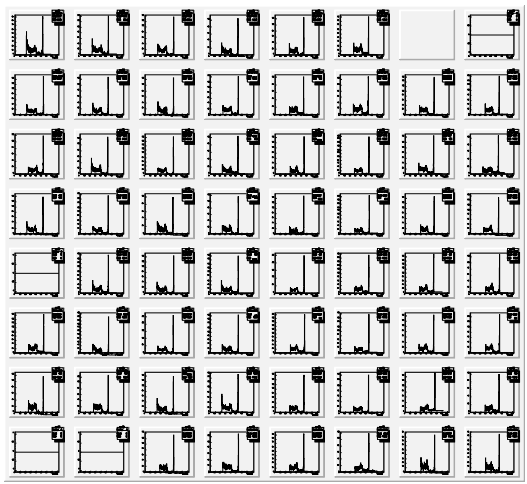

Fig. 4(a) 662 keV($^{137}$Cs source) energy spectra of 60 pixel on a $1\times2\times2$ cm$^3$ CZT detector. 59 out of 60 pixels show proper signals.

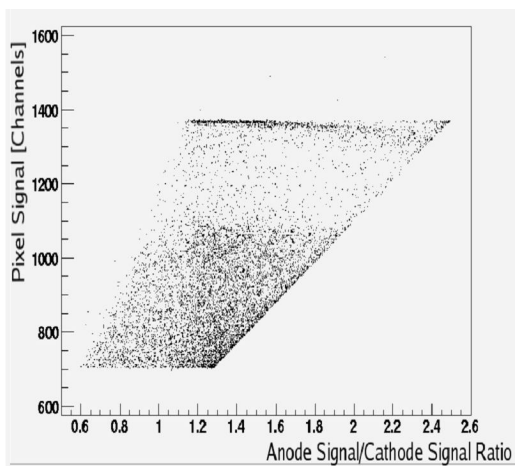

Fig. 4(b) Correlation of the anode charge versus the anode-to-cathode charge ratio for one pixel of a large-volume ($1\times2\times2$ cm$^3$) CZT detector from the company eV-Products (64 pixels, pixel pitch 2.5 mm).

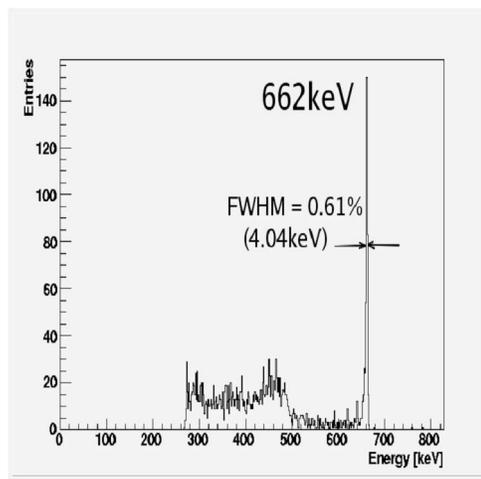

Fig. 4(c) 662 keV energy spectrum of one pixel of a large-volume ($1\times2\times2$ cm$^3$) CZT detector from the company eV-Products after DOI correction (64 pixels, pixel pitch 2.5 mm). Averaged over all pixels of the detector, the energy resolution is 0.79% (5.23 keV).

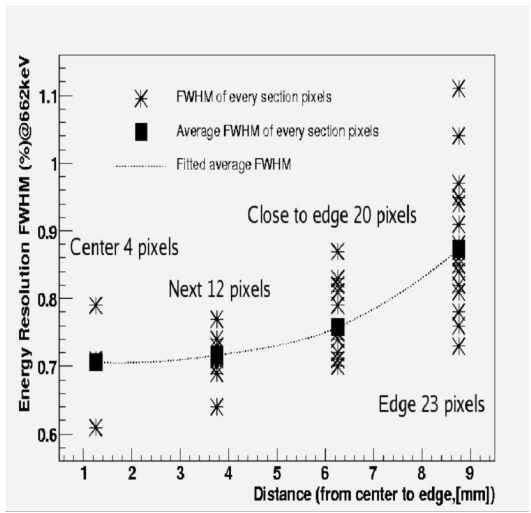

Fig. 4(d) Dependence of the energy resolution on the distance of the pixels from the center of the detector for the same detector as Fig. 4(a)(b) and (c).

## C. Test of detectors with different pixel pitches and different thicknesses

Fig. 5(a) and 5(b) show 662 keV energy spectra obtained with detectors (all: 0.5×2×2 cm$^3$) contacted with pixels at different pixel pitches. An eV-Products detector contacted with 64 pixels at a pitch of 2.5 mm gives energy resolutions of 1.64% (10.86 keV). An eV-Product detector contacted with 121 pixels at a pitch of 1.8 mm gives energy resolutions of 1.08% (7.15 keV). Energy resolutions obtained with detectors of different thicknesses are tabulated in Table 1. The energy resolutions improve with the thickness of the detectors: 1.28% for a 0.5 cm thick Orbotech detector (Fig. 5(c)), 0.73% for a 0.75 cm thick Orbotech detector (Fig. 5(d)), and 0.61% for a 1 cm thick eV-Products detector (Fig. 4(c)). Both results, the energy resolution as function of pixel pitch, and the energy resolution as function of detector thickness indicate that the energy resolution improves with the detector thickness over pixel pitch ratio. The larger the aspect ratio, the better works the small-pixel effect. Fig. 5(e) is energy resolution vs. different thickness of the CZT detectors at 662 keV. We see the same behavior in detector simulations [16].

TABLE 1 Energy spectra resolution and $\mu_e\tau_e$-product results of different thickness CZT detectors at 662 keV

| CZT Thickness (cm) | 64 pixels | Energy Spectra Resolution @662keV | | $\mu_e\tau_e$ (cm$^2$ V$^{-1}$) |
|---|---|---|---|---|
| | | FWHM (%) | FWHM (keV) | |
| 0.5 (Orbotech) | Min. | 1.28 | 8.47 | $4.3 \times 10^{-3}$ |
| | Max. | 2.04 | 13.5 | |
| | Ave. | 1.53 | 10.1 | |
| 0.75 (Orbotech) | Min. | 0.73 | 4.83 | $3.6 \times 10^{-3}$ |
| | Max. | 1.40 | 9.27 | |
| | Ave. | 0.92 | 6.09 | |
| 1 (eV-products) | Min. | 0.61 | 4.04 | $2.1 \times 10^{-2}$ |
| | Max. | 1.11 | 7.35 | |
| | Ave. | 0.79 | 5.23 | |

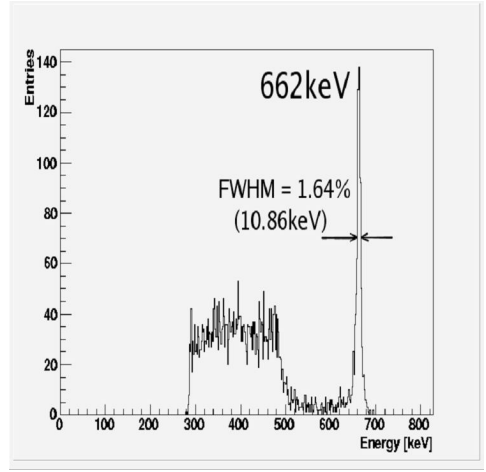

Fig. 5(a) Energy spectrum of a 0.5×2×2 cm$^3$ CZT detector from the company eV-Products contacted with 8×8 pixels (cathode bias: -1100 V).

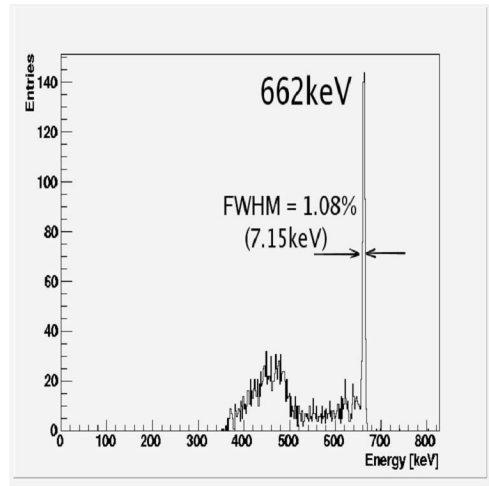

Fig. 5(b) Energy spectrum of 0.5×2×2 cm$^3$ eV products CZT detector with 11×11 pixel (cathode bias voltage: -2000 V).

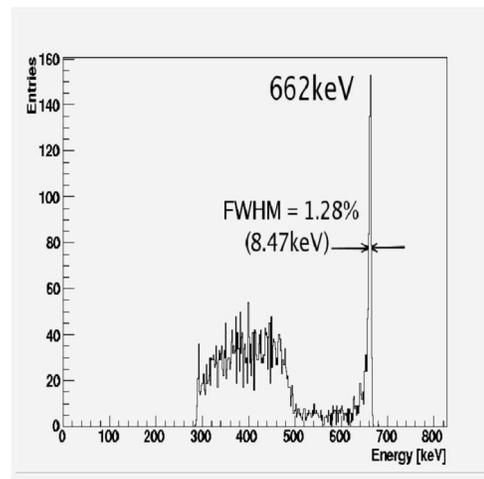

Fig. 5(c) Energy spectrum of 0.5×2×2 cm$^3$ Orbotech CZT detector contacted with 8×8 pixel (cathode bias: -1100 V).

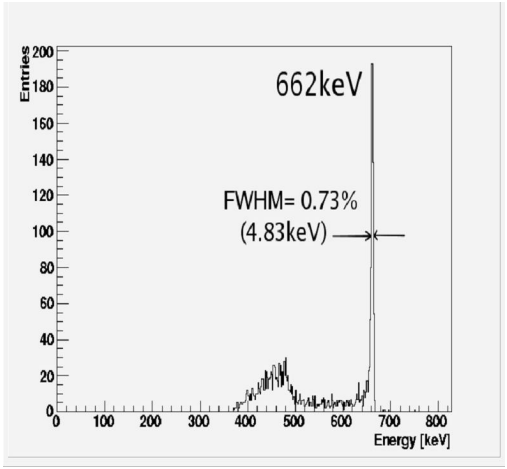

Fig. 5(d) Energy spectrum of a 0.75×2×2 cm³ Orbotech CZT detector contacted with 8×8 pixel (cathode bias: -2000 V).

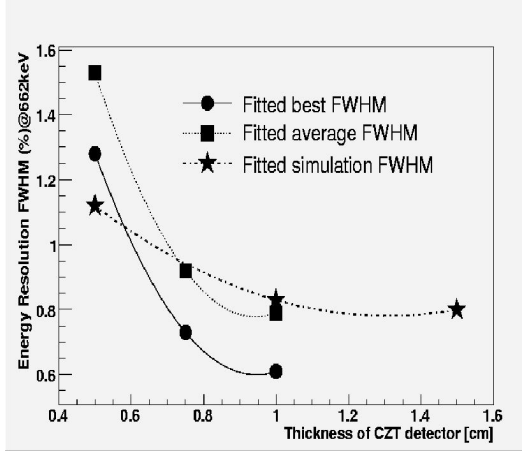

Fig. 5(e) 662 keV energy resolution vs. the thickness of the CZT detectors. The simulation results are from the literature [16].

Tendency can be understood if one takes into account that the energy resolution deteriorates "close" to the pixel where the closeness is the distance from the pixels measured in units of the pixel pitch. Close to the pixels, little charge is induced on the cathode, deteriorating the DOI resolution. Furthermore, close to a pixel, the induced charge depends strongly on the exact location of the energy deposition, and the DOI correction is less effective.

D. *Electron Mobility- Lifetime Products*

The electron mobility-lifetime product can be calculated with the equation:

$$\mu_e \tau_e = \frac{D^2}{\ln\left(\frac{H_{a1}}{H_{a2}}\right)}\left(\frac{1}{V_2} - \frac{1}{V_1}\right) \quad (1)$$

where $D$ is the detector thickness, $H_{a1}$ and $H_{a2}$ are the photopeak centroids under two different cathode biased voltages $V_1$ and $V_2$ [17]. A energy source ($^{137}$Cs) has to be used to ensure that most of the interaction occurs near the cathode, and that hole contributions can be neglected. Maps of $\mu_e\tau_e$-values measured in this way for all the pixels with different CZT detectors are shown in Fig. 6(a), 6(b) and 6(c).

The $\mu_e\tau_e$-products of HPB CZT substrates are given by Zhang et al. [9]. Zhang et al. [9] reported the $\mu_e\tau_e$-product of two different HPB CZT substrates with averages of 3.22×10⁻³ cm² V⁻¹ and of 6.79×10⁻³ cm² V⁻¹ and RMS values of 0.22×10⁻³ cm² V⁻¹ and 0.51×10⁻³ cm² V⁻¹, respectively. Our measured $\mu_e\tau_e$-product results are 4.4×10⁻³ cm²V⁻¹, 3.6×10⁻³ cm²V⁻¹ and 2.1×10⁻² cm²V⁻¹ and RMS values are 1.4×10⁻³ cm²V⁻¹, 0.52×10⁻³ cm² V⁻¹ and 0.44×10⁻² cm² V⁻¹ for 0.5 cm, 0.75cm and 1 cm CZT substrates, respectively.

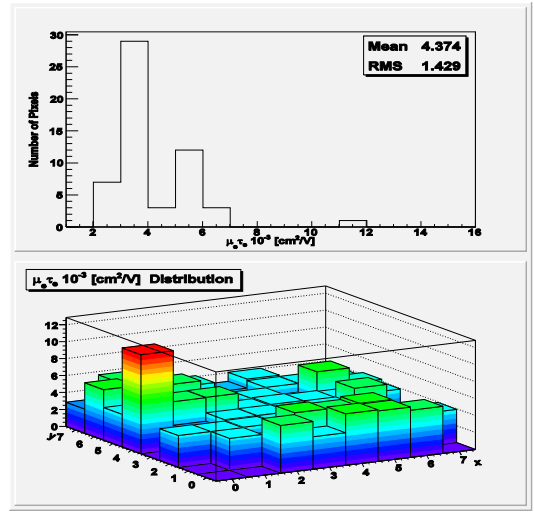

Fig. 6 (a) Electron mobility-lifetime product of 0.5×2×2 cm³ CZT detector with 8×8 pixel (Orbotech).

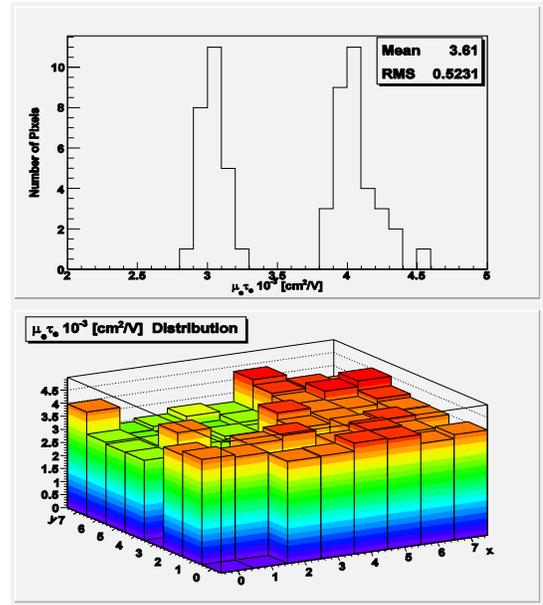

Fig. 6 (b) Electron mobility-lifetime product of 0.75×2×2 cm³ CZT detector with 8×8 pixel (Orbotech).

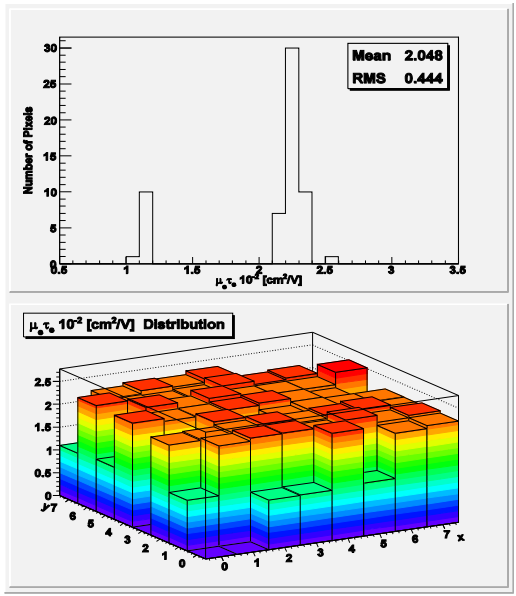

Fig. 6 (c) Electron mobility-lifetime product of 1×2×2 cm³ CZT detector with 8×8 pixel (eV-Products).

## IV. SUMMARY AND OUTLOOK

In this contribution we have described a low-noise ASIC-based readout system for the test of CZT detectors, and have shown first results obtained with detectors of different thicknesses, contacted with pixels at different pitches.

We obtained the following important results:
• The ASIC readout achieves a low level of readout noise. The CZT noise equivalent is 1.5 keV FWHM.
• We obtained <1% energy resolutions with an 0.75×2×2 cm³ CZT substrate from the company Orbotech, and with a 1×2×2 cm³ CZT substrate from the company eV-Products. This is the first time that <1% energy resolutions have been reported for CZT detectors of the company Orbotech.
• We see a trend that the energy resolution improves for larger detector-thickness over pixel-pitch ratios. Our result was obtained with a considerable number of different substrates.

The dependence of the energy resolution on the detector-thickness over pixel-pitch ratio can also be studied by contacting the same substrates with different pixel patterns. We have started such a study and the results will be reported in a forthcoming paper.

ACKNOWLEDGEMENTS
This work is supported by NASA under contract NNX07AH37G, and the DHS under contract 2007DN077ER0002.